\documentclass[journal]{IEEEtran}

\usepackage{xcolor}

\usepackage{algorithmic}

\usepackage{color}

\usepackage{cite}

\ifCLASSINFOpdf
  \usepackage[pdftex]{graphicx}
 
\else

\fi

\usepackage{algorithmic}
\usepackage{algorithm}

\usepackage{epstopdf,float,amsfonts}
\usepackage{amsmath}
\usepackage{multirow}
\usepackage{multicol}   
\usepackage{float}

\usepackage{footnote}
\makesavenoteenv{tabular}

\usepackage{booktabs}
\usepackage{caption}
\usepackage{subcaption}

\usepackage{caption}

\usepackage{fixltx2e}

\begin{document}

\title{Modeling and Quantifying the Impact of Wind Power Penetration on Power System Coherency}

\author{Sayak Mukherjee$^{1}$,~\IEEEmembership{Student Member,~IEEE,} Aranya Chakrabortty$^{1}$, ~\IEEEmembership{Senior Member,~IEEE,} and Saman Babaei$^{2}$, ~\IEEEmembership{Member,~IEEE}
\thanks{$^{1}$ S. Mukherjee and A. Chakrabortty are with the Electrical and Computer Engineering Department, North Carolina State University, Raleigh 
NC, USA. Emails:\{smukher8, achakra2\}@ncsu.edu.\par
$^{2}$ S. Babaei is currently with Arrivo Inc., CA, USA. During this research he was with the New York Power Authority, White Plains, NY.  Email: sbabaei@ncsu.edu.
}}

\maketitle

\begin{abstract}
This paper presents a mathematical analysis of how wind generation impacts the coherency property of power
systems. Coherency arises from time-scale separation in the dynamics of synchronous generators, where
generator states inside a coherent area synchronize over a fast time-scale due to stronger coupling, while the areas themselves synchronize over a slower time-scale due to weaker coupling. This time-scale separation is reflected in the form of a spectral separation in the weighted Laplacian matrix describing the swing dynamics of the generators. However, when wind farms with doubly-fed induction generators (DFIG) are integrated in the system then this Laplacian matrix changes based on both the level of wind penetration and the location of the wind farms. The modified Laplacian changes the effective slow eigenspace of the generators. Depending on penetration level, this change may result in changing the identities of the coherent areas. We develop a theoretical framework to quantify this modification, and validate our results with numerical simulations of the IEEE $68$-bus system with one and multiple wind farms. We compare our model-based results on clustering with results using measurement-based principal component analysis to substantiate our derivations.
\end{abstract}


\IEEEpeerreviewmaketitle
\vspace{-.3 cm}
\section{Introduction}
\IEEEPARstart{O}{ver} the past two decades, significant amount of research has been done in studying various impacts of wind power penetration on power system dynamics, stability, and control. Results reported in \cite{eknath,Gautam09,Sauer}, for example, demonstrate the impact of wind integration on transient stability and small-signal stability. Results in \cite{goran} show the low-inertial effects of wind integration resulting in degradation of frequency control. Results in \cite{Slootweg2003} overview the effects of  penetration level of wind farms on the damping characteristics of inter-area oscillations. Multiple studies have also been done on developing new control schemes for improving the steady-state and dynamic operation \cite{ajjarapu}, and damping of inter-area oscillations in wind-integrated power systems \cite{Tsourakis2009, miao2009, wind4}. In \cite{Souvik_ts}, an analytical framework has been developed to evaluate the effect of wind penetration on time-scale separation properties of power systems that lead to inter-area oscillations. The majority of these works describe how wind power impacts the eigenvalues of the small-signal model of a power system. However, in order to understand its holistic effect, it is equally important to evaluate how the eigenvectors of these models change as more DFIGs penetrate the grid. 
\par
In this paper we address this problem by studying the impact of wind penetration on a specific eigen-property of power systems that requires the use of eigenvectors - namely, coherency \cite{chowpower},\cite{chowbook}. Coherency is a fundamental property of power systems that arise from  the separation of time-scales in the dynamics of synchronous machines. Machines that are strongly coupled tend to swing together, and synchronize over a fast time-scale, thereby forming {\it coherent} groups or clusters, while the groups themselves swing against each other and synchronize over a slow time-scale due to their weaker coupling. A large literature exists on coherency theory, for example see \cite{podmore,verghese,Wu,bullo}, which to date is still highly useful for transmission planning and operations including dynamic equivalencing \cite{dyneqvittal}, controlled islanding \cite{island}, and oscillation damping control \cite{app_wac}. Recent papers such as  \cite{non-sync, coh_wind} have used data-driven techniques to evaluate how the traditional notions of coherency are affected by non-synchronous power generation  from wind, but no theoretical reasonings have been presented. The results derived in this paper compensate for this gap by inferring that the intrusion of DFIGs can be viewed as an addition of heterogeneity to the homogeneous dynamics of synchronous machines. This heterogeneity changes the \textit{effective} dynamic coupling between the  synchronous machines, and thereby perturbs the eigenvalues and eigenvectors of the swing dynamics such that the coherent groups change. The change depends on the amount of wind power injected, and the bus locations of the wind farms.

The main contributions of the work can be summarized as follows. We compare electro-mechanical models of multi-machine power systems both with and without wind injection, and quantify the perturbation caused by this injection in the weighted Laplacian matrix associated with the swing dynamics of the synchronous generators. The quantification is done in terms of both penetration level and location of wind plants. Using singular perturbation theory, an \textit{equivalent} Laplacian matrix is derived to capture the modified interaction between the synchronous generators in presence of wind. Thereafter, a coherency grouping algorithm is stated in terms of the eigenvectors of this modified Laplacian matrix.  A motivational example for the study is drawn from the model of the New York state power grid with large-scale wind penetration. The theoretical analyses and algorithms are all verified using simulations of the IEEE benchmark 16-machine, 68-bus power system model with multiple wind plants.

\section{Recapitulation of Coherency}
We first recall the fundamental theory of coherency in synchronous machines, as detailed in \cite{chowpower}. Consider a power system with $m$ buses and $n$ synchronous generators. Considering classical model of synchronous generators \cite{kundur}, the dynamics of the $i^{th}$ generator can be written as, 
\vspace{-.2 cm}
\begin{align}\label{eqn:swing}
\hspace{-.3 cm} {{{\dot \delta }_i}}&={{\omega _i}},\;
 M_{i} \dot{\omega} _i = P_{mi}-\frac{E_{i}}{x_{di}^{'}}\left( {{V_{i_{Re}}}\sin {\delta _i} - {V_{i_{Im}}}\cos {\delta _i}} \right),
\end{align}
where $\delta_i$, $\omega_i$, $M_{i}$, $x'_{di}$, $E_i$, $P_{mi}$ are respectively the phase angle, machine speed deviation from nominal speed ($120\pi$ rad/s), inertia, direct-axis transient reactance, internal machine voltage, and the mechanical power input to generator $i \; (i=1,\dots,n)$. $V_{i_{Re}}$ and ${V_{i_{Im}}}$ are the real and imaginary parts of the bus voltage phasor. The active and reactive power outputs of the $i^{th}$  generator can be written as,
\vspace{-.2 cm}
\begin{subequations}
\label{eqn:syn_power}
\begin{align}
{P_{si}} =& \frac{{{E_i}}}{{x_{di}^{'}}}\left( {{V_{{i_{Re}}}}\sin {\delta _i} - {V_{{i_{Im}}}}\cos {\delta _i}} \right),\label{eqn:syncmc_active_i}\\
{Q_{si}} =& \frac{{E_i^2}}{{x_{di}^{'}}} - \frac{{{E_i}}}{{x_{di}^{'}}}\left( {{V_{{i_{Re}}}}\cos {\delta _i} - {V_{{i_{Im}}}}\sin {\delta _i}} \right).\label{eqn:syncmc_reactive_i}
\end{align}
\end{subequations}
The active and reactive power flow balance at any bus $j, j=1,\dots,m$ can be written as,
 \begin{subequations}
\label{eqn:load_flows}
\begin{align}
\label{eqn:load_active} 0=&{P_{ej}} - {\rm{Re}}\left\{
{\sum\limits_{k = 1,k \ne j}^N { {{V_{j}}}{{\left(
{V_{jk}}{B_{jk}} \right)}^*}} } \right\} - V_j^2{G_j},\\
\label{eqn:load_reactive4} 0=&{Q_{ej}} - {\mathop{\rm
Im}\nolimits} \left\{ {\sum\limits_{k = 1,k \ne j}^N
V_{j}{{{\left( {V_{jk}}{B_{jk}}\right)}^*}} }
\right\} - V_j^2{B_j},
\end{align}
\end{subequations}
where $G_j$ and $B_j$ are the conductance and the susceptance of the shunt load at bus $j$ with line charging. Assuming the transmission lines to be lossless, $B_{jk}$ denotes the susceptance of the tie-line connecting bus $j$ and bus $k$. Linearizing \eqref{eqn:swing} and \eqref{eqn:syn_power} about a stable operating point $p_0 = \{\delta_0,0,V_{Re_0},V_{Im_0}\}$ governed by the power flow solution we get,
\begin{subequations}\label{eqn:syncgen_matrixform_nom}
\begin{align}
& \Delta\dot { \delta}  = I\Delta{\omega},\;\;
 M\Delta \dot {\omega}  = {K_{11}}\Delta \delta  + {K_{12}}\Delta V  + \Delta { P_m},\\
\label{eqn:syncmc_active}
&\hspace{-.65 cm} \Delta {P_{s}} =  - {K_{11}}\Delta \delta  - {K_{12}}\Delta V , 
\Delta {Q_{s}} = - {K_{21}}\Delta \delta  - {K_{22}}\Delta V .
\end{align}
\end{subequations}
Here, $M$ is the diagonal matrix of the machine inertias, $K_{ij},i,j=\{1,2\}$ are the Jacobian matrices of appropriate dimensions following from (1)-(3), and $I$ is the identity matrix. $\Delta \delta, \Delta \omega, \Delta P_m, \Delta P_s, \Delta Q_s$ and $\Delta V$ are the vectors constructed by stacking the state, the input and the algebraic variables. Specifically, $\Delta V=[\Delta V_{1_{Re}} \; \dots\; \Delta V_{m_{Re}} \; \Delta V_{1_{Im}} \;\dots \; \Delta V_{m_{Im}}]^T$.  
Following Kron-reduction, and the unforced small-signal model (4) can be written as
\vspace{-.3 cm}
\begin{align}\label{eq_mainmodel}
&\Delta \ddot{\delta} = M^{-1}\underbrace{(K_{11} - K_{12}A_{3}^{-1}A_{1})}_{ \mathcal{L}_0}\Delta\delta,
\end{align}
\vspace{-.5 cm}
\footnotesize
\begin{align*}
&A_{3}= \begin{bmatrix} (-K_{1}-K_{12})^T &  K_{1}^{'T} & (-K_{2}-K_{22})^T  &
K_{2}^{'T} \end{bmatrix}^T \\
&{A_{1}}=\left[ {-K_{11}^T}\;\;\;\,0\;\;{{-K_{21}^T}}\;\,\;\,0\right]^T.
\end{align*}
\normalsize 
The Jacobian matrices $K_{1},K_{2}$ follow from the linearization of \eqref{eqn:load_flows} for the generator buses, while $K'_{1},K'_{2}$ are those for non-generator buses.
\par
We impose a two-time scale behavior on (5) by assuming the network to be divided into $r$ distinct and non-overlapping coherent areas \cite{chowpower}. The two time-scale model can be derived as follows. Let there be $n^\alpha$ generators in area $\alpha, \alpha =1,\dots,r$. Let  $ \Delta\delta_i^{\alpha}$ and $ M_i^{\alpha} $ be the small-signal phase angle and inertia of  the $i^{th}$ machine in area $\alpha$. Define two variables $q_s^\alpha \in \mathbb{R}, q_{f}^\alpha =\mbox{col}(q_{fj}^\alpha) \in \mathbb{R}^{n_\alpha - 1}$ for $\alpha = 1,\dots,r, j=2,\dots,n_{\alpha}$ as,
\begin{align}
 &q_s^{\alpha} = \frac{\sum_{i=1}^{n_{\alpha}} M_i^{\alpha}\Delta\delta_i^{\alpha}}{M^{\alpha}},\;\; q_{f{j}}^{\alpha} =  \Delta\delta_j^{\alpha} - \Delta\delta_1^{\alpha},  
\end{align}
where $M^{\alpha} = \sum_{i=1}^{n_{\alpha}}M_i^{\alpha}$. Stacking $q_s^{\alpha}$ and $q_f^{\alpha}$ into vectors for $\alpha =1,\dots,r$, one can write,
\begin{equation}\label{similarity}
\left[ \begin{array}{c} q_s\\ q_f \end{array} \right] = \begin{bmatrix}  \hat{M}^{-1}U^{T}M \\ G   \end{bmatrix}  \Delta\delta := \begin{bmatrix}  C \\ G   \end{bmatrix}  \Delta\delta,
\end{equation}
where, $\hat{M} = diag(M^1,M^2,..,M^r)$, $U=blockdiag(U_1,U_2,..,U_r)$ with $U_{\alpha} \in \mathbb{R}^{n_\alpha}$ being the vector of all ones, $G =blockdiag(G_1,G_2,..,G_r) $ where definition of $G_{\alpha}$ can be found in \cite{chowpower}.
The transformation (7) is invertible with the inverse given by $[ U \   \ G^{T}(GG^{T})^{-1}]$. 
Using \eqref{similarity}, one can rewrite \eqref{eq_mainmodel} in the time-scale separated form,
\begin{align}
\label{ps_timescales}
\begin{bmatrix} \ddot{q}_s \\ \ddot{q}_f \end{bmatrix} = \begin{bmatrix} T_{11} & T_{12} \\ T_{21} & T_{22}   \end{bmatrix}\begin{bmatrix} q_s \\ q_f \end{bmatrix} := T\begin{bmatrix} q_s \\ q_f \end{bmatrix} ,
\end{align}
\footnotesize
\begin{align}
&T_{11} = \epsilon C M^{-1}\mathcal{L}_0^EU,
T_{12} =  \epsilon C M^{-1}\mathcal{L}_0^E G^{\dagger},
T_{21} = \epsilon G M^{-1}\mathcal{L}_0^EU, \\
&T_{22} = GM^{-1} (\mathcal{L}_0^I) G^{\dagger}   +  \epsilon G M^{-1}\mathcal{L}_0^E G^{\dagger}, 
\end{align}
\normalsize
where $G^{\dagger} = G^T(GG^T)^{-1}$, the matrices $\mathcal{L}_0^I, \mathcal{L}_0^E$ follow from partitioning $\mathcal{L}_0$ as 
\begin{align}
    \mathcal{L}_0 = \mathcal{L}_0^I + \epsilon \mathcal{L}_0^E,
\end{align}
and $0<\epsilon \ll 1$ is a singular perturbation parameter arising from the worst-case ratio of the tie-line reactances internal and external to the coherent areas. For precise definition of $\epsilon$, please see \cite{chowpower}. The model \eqref{ps_timescales} is in the singularly perturbed form where $q_s, q_f$ are the slow and fast variables. For small values of $\epsilon$, \eqref{eq_mainmodel} will exhibit a two time-scale behavior, reflected through one DC mode, $n-r$ fast oscillation modes, and $r-1$ slow oscillation modes. The $r$ coherent groups can be identified from the eigen-analysis of $M^{-1}\mathcal{L}_0$. Algorithm 1 as in \cite{chowpower} recalls the steps for this identification. 
\begin{algorithm}
\footnotesize
\caption{Coherency Identification}
1. Assuming $M$ and $\mathcal{L}_0$ to be known, compute  $(r-1)$ smallest eigenvalues (in magnitude) of $M^{-1}\mathcal{L}_{0}$. Construct $V$, a matrix whose columns are the eigenvectors of the zero eigenvalue and these eigenvalues of $M^{-1}\mathcal{L}_0$.\\
2. Apply Gaussian elimination with full pivoting to $V$. From the pivots obtain identity of the reference generators.\\
3. Permute the rows of $V$ to form $V_r = \begin{bmatrix} V_{r1}\\V_{r2}  \end{bmatrix}$, where $V_{r1}$ are the rows of $V$ corresponding to the reference machines in order, and $V_{r2}$ are the remaining rows of $V$ in order. \\
4. Construct $L := V_{r2}\times V_{r1}^{-1}$. The rows and columns of $L$ correspond to the indices of generators and areas respectively. Let $i^{*} = \underset{i}{\operatorname{argmax}}\; |L(i,j)|$ for a particular $j$. Then the generator corresponding to the $i^{*th}$ row of $L$ belongs to coherent area $j$.

\end{algorithm} 
\normalsize
\vspace{-.6 cm}
\section{Motivating Example}
We next provide a motivating example from the New York State (NYS) power grid to show how wind penetration can change coherency. The utility-scale model of the NYS grid is simulated using the PSS/E. The model consists of over 70,000 buses, and thousands of dynamic elements. The grid is divided into eleven
zones following the NYISO zonal separation based on the similarity of frequency responses of generators in any zone following contingencies.
In each zone, a representative bus is chosen corresponding to the largest generating unit. The system is excited with different NYISO-specified contingencies. Time responses of the frequencies at these buses are recorded in a matrix $\mathbb{M} \in \mathbb{R}^{\rho \times s}$. For this example  $\rho = 11$ is the number of representative buses, and $s=3571$ is the number of data samples. The simulations were run for $15$ seconds following the contingencies, and a sampling time of $0.0042$ second was used.
\par
Next, Principal Component Analysis (PCA) \cite{pca_balarko} is applied on $\mathbb{M}$ with the objective of expressing this matrix as $\mathbb{M} = \mathcal{K}\mathcal{V}^T$ where columns of $\mathcal{V}$ are orthonormal basis vectors. To achieve this the following steps are applied. \\ 
\textbf{Step 1:} The singular value decomposition of $\mathbb{M}$ is computed as $\mathbb{M} = U\Sigma V^T$ where $U \in \mathbb{R}^{\rho \times \rho}$ and $V \in \mathbb{R}^{s \times s}$ are respectively the matrices of left and right singular vectors, and $\Sigma \in \mathbb{R}^{\rho \times s}$ is the diagonal matrix of singular values.\\
\textbf{Step 2:} Since the columns of $V$ are the normalized right eigenvectors of $\mathbb{M}^T \mathbb{M}$, one can write $\mathcal{K} = U\Sigma, \mathcal{V}^T = V^T$.\\
\textbf{Step 3:} The columns of $\mathcal{K} \in \mathbb{R}^{\rho \times s}$ represent the weighting for each principal component. Since $s \gg \rho$, $\mathbb{M}$ can be expected to be a low rank matrix. The weightings for $c$ dominant principal axes of $\mathbb{M}$ (for any chosen $c$) can be found by identifying the $c$ columns of $\mathcal{K}$ that have highest variance. These $c$ columns of $\mathcal{K}$ are finally plotted in a $c$-dimensional plot.  

\begin{figure}[H]
\centering
\includegraphics[width=.9\linewidth,height= 3 cm]{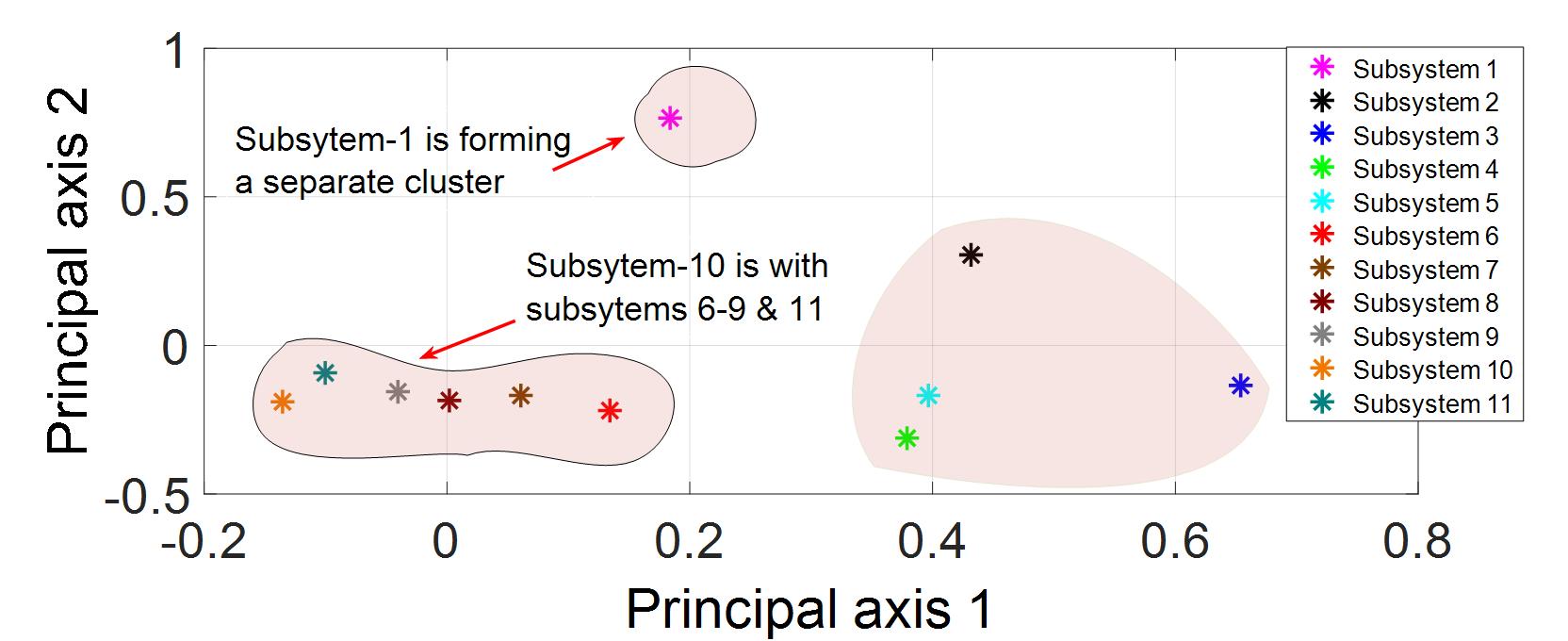}
\caption{\small{PCA weightings for scenario A} }
\label{fig:pca2019}
\end{figure}

\begin{figure}[H]
\centering
\includegraphics[width=.9\linewidth,height= 3 cm]{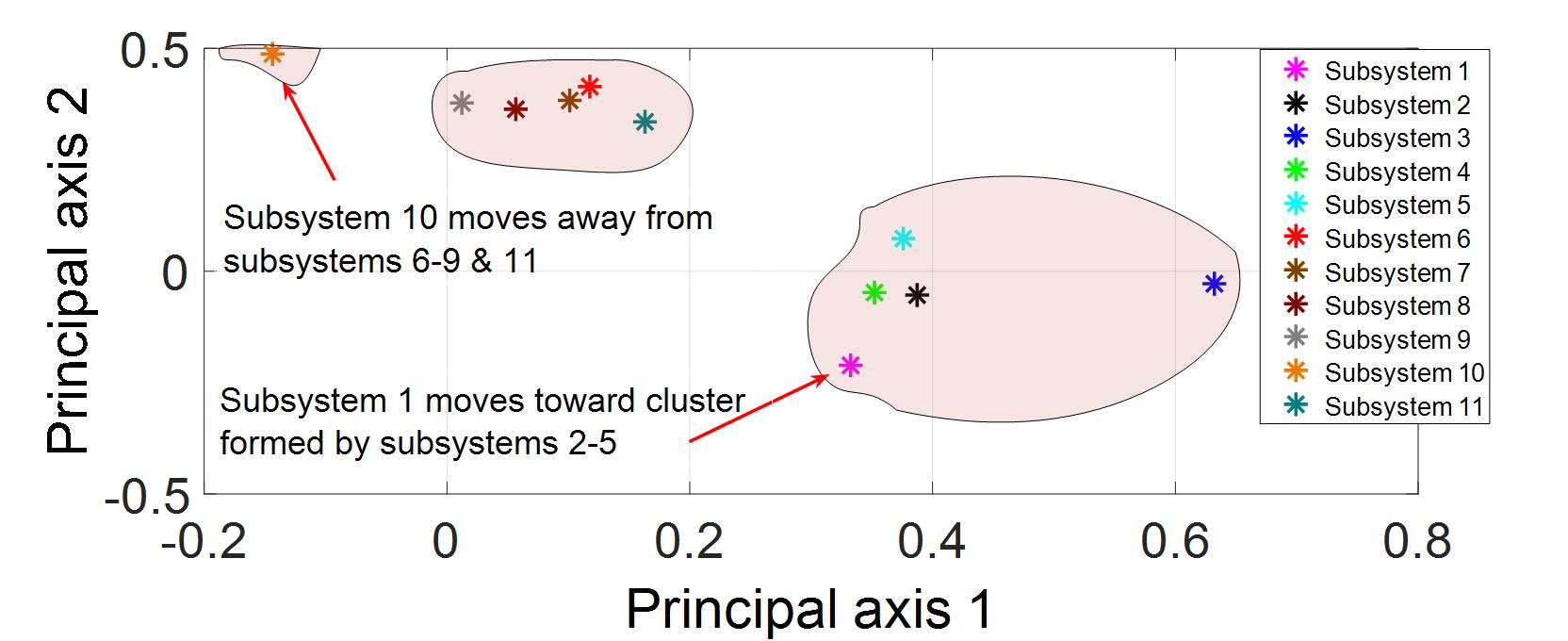}
\caption{\small{PCA weightings for scenario B} }
\label{fig:pca2021}
\end{figure}

For our example we consider two scenarios A and B, where the wind penetration in the NYS model is respectively $10 \%$ and $8 \%$ of the nameplate wind capacity. Most of the wind generation is located in zone $1$ (western NY) and zones $2$-$5$ (central-northern NY). PCA results for these scenarios are shown in Figs. \ref{fig:pca2019}-\ref{fig:pca2021} for the same contingency, considering $c=2$ principal axes. From the two figures it can be seen that subsystem $1$ in scenario A moves towards the group formed by subsystems $2$ through $5$ in scenario B. On the other hand, subsystem $10$ departs from its own coherent group in scenario A, and forms a separate cluster in the scenario B.
\vspace{-.3 cm}
\vspace{-.8 cm}
  This example shows that depending on operating conditions wind penetration can result in changes in coherent clusters. In the next sections we derive conditions that quantify this movement depending on the amount and the location of wind penetration for the power system model \eqref{eq_mainmodel}.
  
\vspace{-.3 cm}
\section{Quantification of perturbation}

\subsection{Wind farm (WF) model}
Consider the $m$-bus power system model introduced in Section II. Without loss of generality we assume that a wind farm is connected at the $(n+1)^{th}$ bus. Following \cite{Souvik_ts}, the farm is assumed to consist of parallel combinations of $\gamma$ individual wind turbine and DFIG units. The turbine model is considered as  
\begin{subequations}\label{eqn:wind_model}
\allowdisplaybreaks
\begin{align}
  J_r\dot{\omega}_r(t)&= \tfrac{B_{dt}}{N_g}\omega_g(t)-K_{dt}\theta_{T}(t)-(B_{dt}+B_r)\omega_r(t)+T_a(t), \label{eqn:wind_model-a}\\
  J_g\dot{\omega}_g(t)&= \tfrac{B_{dt}}{N_g}\omega_r(t)+\tfrac{K_{dt}}{N_g}\theta_{T}(t)
 -\left(\tfrac{B_{dt}}{N_g^2}+B_g\right)\omega_g(t)-T_g(t),\label{eqn:wind_model-b}\\
  \dot\theta_{T}(t)&=\omega_r(t)-\tfrac{1}{N_g}\omega_g(t),\label{eqn:wind_model-c}
\end{align}
\end{subequations}
where ${T_a(t)} = \frac{{\rho {A_s}\nu^3(t){C_p}}}{{2{\omega _r}(t)}}$
is the aerodynamic torque input. The physical meanings of all variables can be found in \cite{Souvik_ts}. 
The DFIG  dynamics can be expressed in a power-invariant synchronously rotating d-q reference frame as
\vspace{-.3 cm}
\begin{align}\label{eqn:volt_cur}
v_{ds} &= R_si_{ds} + \frac{d}{dt}\psi_{ds} - \omega_e\psi_{qs}\nonumber,
v_{qs} = R_si_{qs} + \frac{d}{dt}\psi_{qs} + \omega_e\psi_{ds},\\
v_{dr} &= R_ri_{dr} + \frac{d}{dt}\psi_{dr} - (\omega_e-\omega_{ge})\psi_{qr} ,\nonumber\\
v_{qr} &= R_ri_{qr} + \frac{d}{dt}\psi_{qr} + (\omega_e-\omega_{ge})\psi_{dr},
\end{align}
where expressions for flux linkages and the generated electrical torque are given by
\begin{align}
&\psi_{qs} = L_{ls}i_{qs} + L_{m}(i_{qs} + i_{qr})\nonumber,
\psi_{ds} = L_{ls}i_{ds} + L_{m}(i_{ds} + i_{dr}),\\
&\psi_{qr} = L_{lr}i_{qr} + L_{m}(i_{qs} + i_{qr})\nonumber,
\psi_{dr} = L_{lr}i_{dr} + L_{m}(i_{ds} + i_{dr}),\\
 &{T_g}(t) = \frac{3}{2}\frac{{p}}{2}{L_m}\left[
{{i_{qs}(t)}{i_{dr}(t)} - {i_{ds}(t)}{i_{qr}(t)}} \right].
\end{align}
Here, $\omega_{ge}:=\frac{p_e}{2}\,\omega_{g}$ is the electrical speed of the rotor of the DFIG, $p_e$ is the number of electrical poles. $L_{ls}$, $L_{lr}$, $L_m$ are the stator and rotor leakage inductances and the magnetizing inductance, respectively. Standard meanings of the voltage and flux variables can be found in \cite{Souvik_ts} and are skipped here for brevity. The active and reactive power output of the wind farm can be written as
\begin{subequations}
\label{eqn:im_power}
\begin{align}
 {P_{w}}  = \gamma({v_{qs}}{i_{qs}} +{v_{ds}}{i_{ds}}),
 {Q_{w}} = \gamma(-{v_{ds}}{i_{qs}} +{v_{qs}}{i_{ds}}) .
\end{align}
\end{subequations}
Aligning $v_{qs}$ with the wind bus voltage phasor, we have ${v_{qs}} = |V_{n+1}|$  and ${v_{ds}=0}$. The DFIG is equipped with active and reactive power control loops with PI controllers whose setpoints are computed using Maximum Power Point Tracking (MPPT) and power flow calculations, respectively.

\vspace{-.3 cm}
\subsection{Linearized wind-integrated model}
Since the stator of the DFIG is directly connected to the wind bus through a step-up transformer, the swing states of the synchronous generators will now be dynamically coupled with the wind farm states $z$ which are considered to be the average of the individual unit states \cite{Souvik_ts}.  Let the operating point of the power system after wind integration be denoted as $\hat{p}_0 = \{\hat{\delta}_0,0,z_0,\hat{V}_{Re0},\hat{V}_{Im0}\}$.
Equations \eqref{eqn:swing} and \eqref{eqn:syn_power} are linearized about $\hat{p}_0$  as
\begin{subequations}\label{eqn:syncgen_matrixform1}
\begin{align}
\label{eqn:syncmc_active0}
& \Delta\dot { \delta}  = I\Delta{\omega},\;\;
 M\Delta \dot {\omega}  = {\bar{K}_{11}}\Delta \delta  + {\bar{K}_{12}}\Delta V  + \Delta { P_m},\\
\label{eqn:syncmc_active1}
&\hspace{-.4 cm} \Delta {P_{s}} =  - {\bar{K}_{11}}\Delta \delta  - {\bar{K}_{12}}\Delta V ,\;\;  
\Delta {Q_{s}} = - {\bar{K}_{21}}\Delta \delta  - {\bar{K}_{22}}\Delta V.
\end{align}
\end{subequations}
The Jacobians $\bar{K}_{ij}$ are different than $K_{ij}$ in \eqref{eqn:syncgen_matrixform_nom} due to the shift in operating point from the wind injection. The linearized wind farm model from (12)-(14) around $\hat{p}_0$ is written as
\vspace{-.2 cm}
\begin{equation}
\label{eqn:im_ss}
\Delta {\dot z} = {A}\Delta {z} + {B}\Delta V.
\end{equation}
Expressions for $A$ and $B$ are skipped for brevity. Note that $B$ is a zero padded matrix where zeros correspond to the non-wind buses. The linearised power output equations are written as
\begin{align}
\label{eqn:im_linearpower}
& \hspace{1.3 cm} \left[ {\begin{array}{*{20}{c}}
{{\Delta P_{w}}}\\
{{\Delta Q_{w}}}
\end{array}} \right] = \left[ {\begin{array}{*{20}{c}}
{{C_1}\Delta z}\\
{{C_2}\Delta z}
\end{array}} \right] + \left[ {\begin{array}{*{20}{c}}
{{D_{1}}\Delta V}\\
{{D_{2}} \Delta V}
\end{array}} \right],\\ \nonumber
&\hspace{-.44 cm}C_1= \frac{\partial P_w}{\partial z}\Bigr|_{\hat{p}_0} ,C_2 =\frac{\partial Q_w}{\partial z}\Bigr|_{\hat{p}_0} ,D_1=\frac{\partial P_w}{\partial V}\Bigr|_{\hat{p}_0}, D_2 = \frac{\partial Q_w}{\partial V}\Bigr|_{\hat{p}_0}.
\end{align}
The zero-padded matrices $D_1$ and $D_2$ depend on wind penetration level $\gamma$ and on the location of the wind farm. More detailed structures of $D_1$ and $D_2$ will be shown shortly. The nonlinear power flow equations \eqref{eqn:load_flows} are linearized at $\hat{p}_0$ with Jacobian matrices $\bar{K}_1,\bar{K}_2$ for active and reactive power flows for synchronous generator buses, $\bar{K}_3,\bar{K}_4$ for wind generator bus, and $\bar{K}_5,\bar{K}_6$ for the non-generator buses as
\begin{align}\label{eqn:kron_reduce}
0 = {\bar{A}_1}\Delta \delta  + {\bar{A}_2} \Delta z +{\bar{A}_3} \Delta V,
\end{align}
\vspace{-.8 cm}
\begin{align*}
\bar{A}_1 &= [{-\bar{K}_{11}^T}\;\;0\;\,0\;\;{{-\bar{K}_{21}^T}}\;\,0\;\,0]^T,
\; \bar{A}_2 = [0\;\;{{C_1}}\;\;0\;\;0\;\;{{C_2}}\;\;0]^T,\\
 \bar{A}_3 &=[{\left( {-{\bar{K}_1} - {\bar{K}_{12}}} \right)^T} \;\; {({D_{1}} - {\bar{K}_3})}\;\;{{\bar{K}_5^T}}\;\;{\left( {-{\bar{K}_2} - {\bar{K}_{22}}} \right)^T}\, \\  &{({D_{2}} - {\bar{K}_4})}\;\;{{\bar{K}_6^T}} ]^T.
 \end{align*}
Using \eqref{eqn:syncmc_active0}, \eqref{eqn:im_ss} and \eqref{eqn:kron_reduce}, considering $\Delta P_m =0$ we write the final set of state-space equations for the unforced wind-integrated power system model:
\begin{align}
\label{eqn:smallsignal_model}
\hspace{-.5 cm}\left[ \begin{array}{c} \Delta\dot\delta \\ \Delta\dot\omega \\  \Delta\dot z \end{array} \right] = \begin{bmatrix} 0 &  I & 0  \\  \mathcal{R}_1 & 0 &\mathcal{R}_2  \\ \mathcal{R}_3 & 0 & \mathcal{R}_4  \end{bmatrix}  \left[ \begin{array}{c}\Delta\delta \\ \Delta\omega \\  \Delta z  \end{array} \right] ,
\end{align}
where, $\mathcal{R}_1=M^{-1}(\bar{K}_{11} - \bar{K}_{12}\bar{A}_3^{-1}\bar{A}_1)$, $\mathcal{R}_2 = M^{-1}(-\bar{K}_{12}\bar{A}_3^{-1}\bar{A}_2)$,$\mathcal{R}_3=-B\bar{A}_3^{-1}\bar{A}_1$, $\mathcal{R}_4=A-B\bar{A}_3^{-1}\bar{A}_2$. 
\par
We next quantify the perturbation of the wind-integrated model \eqref{eqn:smallsignal_model} from the wind-less model (5). Note that $\mathcal{R}_1=
M^{-1}\mathcal{L}$ where $\mathcal{L} = (\bar{K}_{11} - \bar{K}_{12}\bar{A}_3^{-1}\bar{A}_1)
$ mimics the role of $\mathcal{L}_0$ in (5). However, unlike $\mathcal{L}_0$, which is a weighted Laplacian matrix (by definition, a symmetric matrix $\mathcal{L}_0 = \mathcal{L}_0^T$ is Laplacian if the diagonal entry of each row equals to the negative sum of the other entries of that row), $\mathcal{L}$ is not a Laplacian matrix. To quantify the difference between $\mathcal{L}_0$ and $\mathcal{L}$, we must compare each of their constituent matrices separately. This is shown as follows.


\vspace{-.5 cm}
\subsection{Quantification of perturbation in $\mathcal{L}$}



\begin{itemize}
    \item \textbf{Perturbation in $K_{11}$ and $K_{12}$:}  \\
    The nominal Jacobian matrices $K_{11}, K_{12}$ are perturbed because of the shift in operating point from $p_0$ to $\hat{p}_0$. We write this perturbation as
    \begin{align}\label{K11_delta}
        \bar{K}_{11} = K_{11} + \Delta_{k11}(\gamma), \bar{K}_{12} = K_{12} + \Delta_{k12}(\gamma). 
    \end{align}
The unstructured perturbations $\Delta_{k11}(\gamma), \Delta_{k12}(\gamma)$ are implicit functions of $\gamma$ and the location of wind farm, and can be determined numerically for a particular wind integration scenario.
\item \textbf{Perturbation in $A_3$:}  \\
The matrices $K'_1$ and $K'_2$ in (5) are rewritten as $K'_1 = [-K_3 \; K_5], K'_2 = [-K_4 \; K_6]$ where $K_3,K_4$ are the Jacobians after linearizing (3) for the nominal model with respect to the $(n+1)^{th}$ bus voltage, and $K_5,K_6$ are that with respect to the voltages of non-generator buses. To compare $A_3$ and $\bar{A}_{3}$ we need to consider the structures of $D_1$ and $D_2$. The entries of $D_1,D_2$ can be partitioned in terms of synchronous generator, wind farm and non-generator buses as follows:  

\footnotesize
\begin{align*}
& \hspace{-.6 cm} D_1 = \bordermatrix{ & &\underset{\downarrow}{(n+1)} & & & \underset{\downarrow}{(n+m+1)}& & \cr
       & \underbrace{0}_{\frac{\text{sync-gens}}{1 \times n}} & \underbrace{\frac{\partial P_w}{\partial V_{(n+1)_{Re}}}\Bigr|_{\hat{p}_0} }_{\frac{\text{wind gen}}{(1 \times 1)}} & \underbrace{0}_{\frac{\text{non-gens}}{1 \times (m-n-1)}} & 0 & \frac{\partial P_w}{\partial V_{(n+1)_{Im}}}\Bigr|_{\hat{p}_0} & 0 },\\
& \hspace{-.6 cm} D_2 = \bordermatrix{ & &\underset{\downarrow}{(n+1)} & & & \underset{\downarrow}{(n+m+1)}& & \cr
       & \underbrace{0}_{\frac{\text{sync-gens}}{(1 \times n)}} & \underbrace{\frac{\partial Q_w}{\partial V_{(n+1)_{Re}}}\Bigr|_{\hat{p}_0} }_{\frac{\text{wind gen}}{(1 \times 1)}} & \underbrace{0}_{\frac{\text{non-gens}}{1 \times (m-n-1)}} & 0 & \frac{\partial Q_w}{\partial V_{(n+1)_{Im}}}\Bigr|_{\hat{p}_0} & 0 }.  
\end{align*}
\normalsize
Simple calculations show that the partial derivatives in the above matrices can be written as $\frac{\partial P_w}{\partial V_{(n+1)_{Re}}}\Bigr|_{\hat{p}_0} :=\gamma \zeta_1, \; \frac{\partial P_w}{\partial V_{(n+1)_{Im}}}\Bigr|_{\hat{p}_0} :=\gamma \zeta_2,\frac{\partial Q_w}{\partial V_{(n+1)_{Re}}}\Bigr|_{\hat{p}_0} :=\gamma \zeta_3 , \; \frac{\partial Q_w}{\partial V_{(n+1)_{Im}}}\Bigr|_{\hat{p}_0} :=\gamma \zeta_4$,
where $\zeta_i$'s are linearization constants depending on the steady-state stator voltage and stator currents of the DFIG. 
Using the structures of $D_1$ and $D_2$, we can write:  
\begin{align}\label{A3pert}
& \bar{A}_3 = A_{3} + \gamma A'_3 + \Delta_{A3}(\gamma),
\end{align}
where the perturbation term $A'_3$ has the following sparse structure,
\vspace{-.5 cm}
\footnotesize
\begin{align*}
\hspace{-.3 cm}  A'_3 = \bordermatrix{ & &\underset{\downarrow}{(n+1)} & & & \underset{\downarrow}{(n+m+1)}& & \cr 
       & \dots & \dots & \dots & \dots & \dots & \dots \cr
     (n+1) \rightarrow  & \dots & \zeta_1 & \dots & \dots & \zeta_2 & \dots \cr
      & \dots & \dots & \dots & \dots & \dots & \dots \cr
     (n+m+1) \rightarrow  & \dots & \zeta_3 & \dots & \dots & \zeta_4 & \dots \cr
     & \dots & \dots & \dots & \dots & \dots & \dots },
\end{align*}
\normalsize
and $\Delta_{A3}(\gamma)$ captures the change in the nominal Jacobians $K_{12},K_{22}$ and $K_i, i=1,\dots,6$ due to the operating point shift.
\item \textbf{Perturbation in $A_1$:} \\
To compare $A_1$ in (5) with $\bar{A}_1$ in \eqref{eqn:kron_reduce}, we use \eqref{K11_delta}. We write the perturbation as
\begin{align}\label{A1_pert}
    \bar{A}_1 = A_1 + \Delta_{A1}(\gamma),
\end{align}
where $\Delta_{A1}(\gamma) = [\Delta_{k11}^T\;0\;0\;\Delta_{k12}^T\;0\;0]^T$ follows from (21).
\item \textbf{Perturbation in $\mathcal{L}_0$:}  \\
We recall from (20) and (5) that $\mathcal{L} = \bar{K}_{11} - \bar{K}_{12}\bar{A}_3^{-1}\bar{A}_1$ and $\mathcal{L}_0 = K_{11} - K_{12}A_3^{-1}A_1$. Using the matrix inversion lemma we get,
\begin{align}
 \bar{A}_3^{-1} = (A_{3}+ \gamma A'_3 + \Delta_{A3}(\gamma))^{-1} = A_{3}^{-1} + X,
\end{align}
where $
  X = 
-(I+A_{3}^{-1}(\gamma A'_3 + \Delta_{A3}(\gamma)))^{-1}A_{3}^{-1}(\gamma A'_3 + \Delta_{A3}(\gamma))A_{3}^{-1}
$. Using \eqref{K11_delta}-\eqref{A1_pert} we get,
\begin{align}
 &\hspace{-.5 cm}   \mathcal{L} = ((K_{11}+\Delta_{k11}) - (K_{12}+\Delta_{k12})(A_{3}^{-1}+ X)(A_{1}+\Delta_{A1})), \nonumber \\
 &= \mathcal{L}_{0} \underbrace{-K_{12}XA_{1} +\kappa_{\mathcal{L}}(\gamma)}_{\text{Perturbation: }\Delta \mathcal{L}_0}.\label{pertL}
\end{align}
Here $\kappa_{\mathcal{L}}(\gamma) = \Delta_{k11} - K_{12}\Delta_{A1} - \Delta_{k12}(A_{3}^{-1}+ X)(A_{1}+\Delta_{A1}))$, contains all the terms due to change in operating point, while the perturbation term $-K_{12}XA_{1}$ contains explicit information about penetration level and location of the wind farm. 
\end{itemize}
\vspace{-.34 cm}
\subsection{Extension to multiple wind farms}
Equations (21)-(25) provide the expressions for matrix perturbations when there is only one wind farm in the system at the $(n+1)^{th}$ bus. The approach can be easily extended to when the system has $p$ wind farms, say at $(n+1,n+2,\dots,n+p)^{th}$ buses. Let their penetration levels be $\gamma _1,\gamma _2,\dots,\gamma _p$, respectively. We consider the matrix $D_1 = [D_1^{Re}, D_1^{Im}]$, where $D_1^{Re} \in \mathbb{R}^{p \times m} $ is given by  

\footnotesize
\begin{align}\label{D1mult}
\hspace{-.3 cm} D_1^{Re} = \bordermatrix{& &\underset{\downarrow}{(n+1)}  &\underset{\downarrow}{(n+2)} & & \underset{\downarrow}{(n+p)}&  &\cr 
       & \dots & \frac{\partial P_w}{\partial V_{(n+1)_{Re}}} & \dots & \dots & \dots & \dots\cr
      & \dots & \dots &  \frac{\partial P_w}{\partial V_{(n+2)_{Re}}} & \dots & \dots & \dots \cr
      & \dots & \dots & \dots & \dots & \dots & \dots \cr
       & \dots & \dots & \dots & \dots &  \frac{\partial P_w}{\partial V_{(n+p)_{Re}}} & \dots & }.
\end{align}
\normalsize
The definition of $D_1^{Im}$ is the same as (26) but with $V_{i_{Re}}$ replaced by $V_{i_{Im}},i=n+1,\dots,n+p$. Similarly, $D_2 = [D_2^{Re},D_2^{Im}]$ can be defined in the same way by replacing $P_w$ by $Q_w$ in (26). As before, the partial derivatives are written as $\frac{\partial P_w}{\partial V_{i_{Re}}} = \gamma_i \zeta_{1i}, \frac{\partial P_w}{\partial V_{i_{Im}}} = \gamma_i \zeta_{2i}, \frac{\partial Q_w}{\partial V_{i_{Re}}} = \gamma_i \zeta_{3i},\frac{\partial Q_w}{\partial V_{i_{Im}}} = \gamma_i \zeta_{4i}, i=n+1,\dots,n+p$.  Accordingly, the admittance matrix $\bar{A}_3$ is expressed as
\begin{align}
\bar{A}_3 = A_{3} + \gamma_1 A_3^1 + \gamma_2 A_3^2 + \dots + \gamma_p A_3^p + \Delta_{A3}(\gamma),
\end{align}
where $A_3^i$ has the same structure as $A'_3$ in (22) but with $\zeta_1,\zeta_2,\zeta_3,\zeta_4$ replaced by $\zeta_{1i},\zeta_{2i},\zeta_{3i},\zeta_{4i}$, respectively. The comparison between $\mathcal{L}$ and $\mathcal{L}_0$ follows thereafter in the same way as in the foregoing subsection. 
\section{Perturbation analysis for two time-scale property}
We recall from (11), that $\mathcal{L}_0 = \mathcal{L}_0^I + \epsilon \mathcal{L}_0^E$. We next analyze how the perturbation in $\mathcal{L}_0$ as given by \eqref{pertL} extends to its the internal and external connection components $\mathcal{L}_0^I$ and $\mathcal{L}_0^E$, respectively.
\vspace{-.5 cm}
\subsection{Perturbation in $\mathcal{L}_0^I$ and $\mathcal{L}_0^E$ }

Due to the existence of $r$ clusters one can separate $A_{3}$ into internal and external admittance matrices  as,
$
A_{3}=A_{3}^I + \epsilon A_{3}^E 
$. Using exponential expansion we can write, 
\begin{align}
 A_{3}^{-1} = (A_{3}^I + \epsilon A_{3}^E)^{-1} 
 = (A_{3}^I)^{-1}+\epsilon X_{1\epsilon}, 
\end{align}
where $ X_{1\epsilon} = [(-(A_{3}^I)^{-1}(A_{3}^E) +\epsilon((A_{3}^I)^{-1}(A_{3}^E))^2 - \dots )(A_{3}^I)^{-1}]$. Considering a single wind farm scenario we recall \eqref{pertL} as
\begin{align}
\mathcal{L} = \mathcal{L}^I_0 + \epsilon \mathcal{L}^E_0 -K_{12}XA_{1} + \kappa_{\mathcal{L}}(\gamma) .
\end{align}  
The matrix $X$ can be written as 
\begin{align}
    X= -(I+A_{3}^{-1}x)^{-1}A_{3}^{-1}xA_{3}^{-1},
\end{align}
where $x= (\gamma A'_3)+\Delta_{A3}$. Expanding $(I+A_{3}^{-1}x)^{-1}$ we get,
\begin{align}
(I+A_{3}^{-1}x)^{-1} = P_{1a}^{-1} + \epsilon X_{2\epsilon}, 
\end{align}    
where $P_{1a} = I + (A_{3}^{I})^{-1}x$, and $X_{2\epsilon}$ has the same structure as $X_{1\epsilon}$ with $A_{3}^I, A_{3}^E$ replaced by $P_{1a}$ and $X_{1\epsilon}x$, respectively. In that case, we have
\begin{align}
X &= - (P_{1a}^{-1} + \epsilon X_{2\epsilon})((A_{3}^I)^{-1}+\epsilon X_{1\epsilon})x((A_{3}^I)^{-1}+\epsilon X_{1\epsilon}) \nonumber,\\
& = -( P_a + \epsilon P_b).
\end{align}
Here, $P_a = P_{1a}^{-1}(A_{3}^I)^{-1}x(A_{3}^I)^{-1}$ and $P_{ b} = X_{2\epsilon}((A_{3}^I)^{-1}+\epsilon X_{1\epsilon})x((A_{3}^I)^{-1}+\epsilon X_{1\epsilon}) + P_{1a}^{-1}X_{1\epsilon}x((A_{3}^I)^{-1}+\epsilon X_{1\epsilon})$. Then finally (25) can be rewritten as 
\begin{align}
\hspace{-.3 cm} \mathcal{L} & = \mathcal{L}^I_0 + \epsilon \mathcal{L}^E_0 + K_{12}P_aA_{1} + \epsilon K_{12}P_bA_{1} + \kappa_{\mathcal{L}}(\gamma),\\
&=(\mathcal{L}_0^I + \Delta_{\mathcal{L}^I} ) + \epsilon (\mathcal{L}_0^E + \Delta_{\mathcal{L}^E}),\label{pertL2}
\end{align} 
where $\Delta_{\mathcal{L}^I} = K_{12}P_aA_{1} + \kappa_{\mathcal{L}}(\gamma), \Delta_{\mathcal{L}^E} = K_{12}P_bA_{1}$. These perturbations, which are functions of $\gamma$ and the wind farm location, quantify the changes in the internal and external components of $\mathcal{L}_0$.  
\vspace{.06 cm}
\vspace{-.5 cm}
\subsection{Extraction of the equivalent Laplacian}
We next substitute the expression \eqref{pertL2} in $\mathcal{R}_1 = M^{-1}\mathcal{L}$ in the wind-integrated model \eqref{eqn:smallsignal_model}. From this model we can write 
\begin{align}
    \Delta \ddot{\delta} = M^{-1}((\mathcal{L}_0^I + \Delta_{\mathcal{L}^I} ) + \epsilon (\mathcal{L}_0^E + \Delta_{\mathcal{L}^E})) + \mathcal{R}_2\Delta z.\label{synconly}
\end{align}
Our intent is to capture the interactions between the  synchronous machines in the wind-integrated system. For that we apply the transformation (7) on \eqref{synconly}, which results in the following transformed unforced dynamics 

\vspace{-.3 cm}
\begin{align}
\label{wind_timescales}
\begin{bmatrix} \ddot{\tilde{q}}_s \\ \ddot{\tilde{q}}_f \end{bmatrix} = \underbrace{\begin{bmatrix} \tilde{T}_{11} & \tilde{T}_{12} \\ \tilde{T}_{21} & \tilde{T}_{22}   \end{bmatrix}}_{\tilde{T}}\begin{bmatrix} \tilde{q}_s \\ \tilde{q}_f \end{bmatrix}+ \begin{bmatrix} C\mathcal{R}_2 \\ G\mathcal{R}_2 \end{bmatrix} \Delta z,
\end{align}
where,
\footnotesize
\begin{align}\label{36}
&\tilde{T}_{11} = CM^{-1}\Delta_{\mathcal{L}^I} U + \epsilon C M^{-1}\mathcal{L}_0^EU + \epsilon C M^{-1}\Delta_{\mathcal{L}^E} U\\
&\tilde{T}_{12} = CM^{-1}\Delta_{\mathcal{L}^I} G^{+} + \epsilon C M^{-1}\mathcal{L}_0^E G^{+} + \epsilon C M^{-1}\Delta_{\mathcal{L}^E} G^{+}\\
&\tilde{T}_{21} = GM^{-1}\Delta_{\mathcal{L}^I} U + \epsilon G M^{-1}\mathcal{L}_0^EU + \epsilon G M^{-1}\Delta_{\mathcal{L}^E} U\\
\begin{split}
&\tilde{T}_{22} = GM^{-1} (\mathcal{L}_0^I) G + GM^{-1}\Delta_{\mathcal{L}^I} G^{+}  +  \epsilon G M^{-1}\mathcal{L}_0^E G^{+} \\
&   \quad \quad \quad  \quad \quad \quad \quad \quad \quad  + \epsilon G M^{-1}\Delta_{\mathcal{L}^E} G^{+}.
\end{split}\label{39}
\end{align}
\normalsize
\noindent Here $\tilde{q}_s, \tilde{q}_f$ are the slow and fast states corresponding to synchronous-only motions. Due to space constraints we skip the derivations of \eqref{36}-\eqref{39}. Comparing the nominal coherency dynamics in \eqref{ps_timescales} with that of the perturbed dynamics in \eqref{wind_timescales}, it is clear that if the matrix $\tilde{T}$ in  \eqref{wind_timescales} still has to reflect the coherency between synchronous generators in the wind-integrated model, then we must redefine $\mathcal{L}$ such that $CM^{-1}\Delta_{\mathcal{L}^I}G^{\dagger} = \hat{M}U^T\Delta_{\mathcal{L}^I}G^{\dagger} = 0$ and $\Delta_{\mathcal{L}^I}U = 0$. This can be done simply by defining an \textit{equivalent} Laplacian matrix in the following way: 
\begin{align}\label{eq_lap}
\mathcal{L}_{eq}(i,j) := \mathcal{L}(i,j), \; \mathcal{L}_{eq}(i,i) := - \sum_{j \neq i} \mathcal{L}(i,j).
\end{align}
This definition allows us to write 
\begin{align}
    \label{Leq}
    \mathcal{L}_{eq} = \mathcal{L}_0^I + \Delta_{ \mathcal{L}_{eq}^I} + \epsilon(\mathcal{L}_0^E + \Delta_{ \mathcal{L}_{eq}^E}),
\end{align}
where now both perturbations $\Delta_{\mathcal{L}_{eq}^{I}}$ and $\Delta_{\mathcal{L}_{eq}^{E}}$ are Laplacian matrices since $\mathcal{L}_0^I$ and $\mathcal{L}_0^E$ are Laplacian matrices by default. This satisfies  $\hat{M}U^T \Delta\mathcal{L}_{eq}^IG^{\dagger} = 0$ and $\Delta\mathcal{L}_{eq}^IU = 0$, making $\tilde{T}$ structurally consistent with $T$.

\begin{figure}[H]
                \centering
                \includegraphics[width=1\linewidth,keepaspectratio,trim={4 24 4 4},clip]{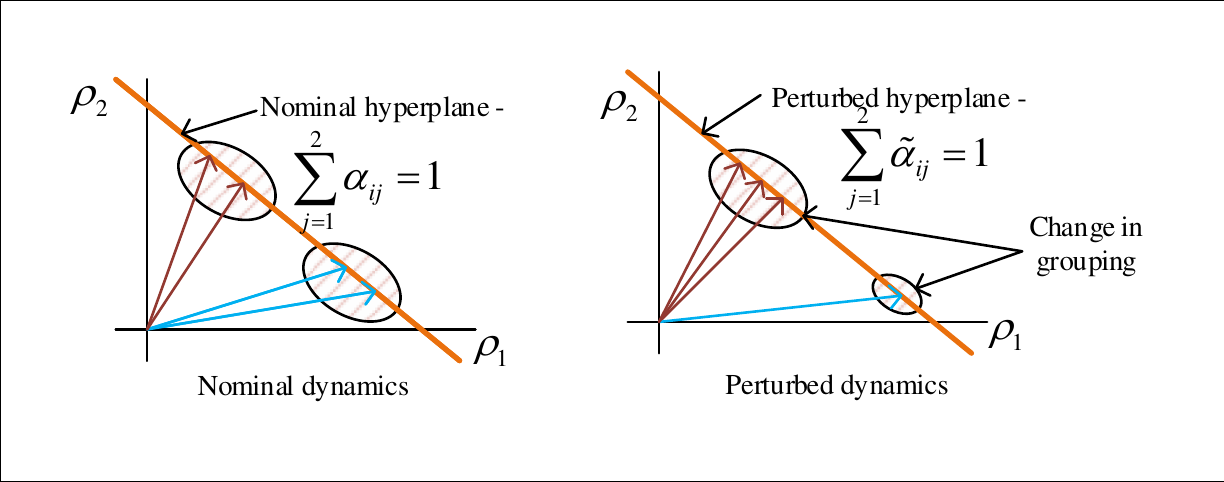}
                \caption{Intuition behind change in grouping}
                \label{fig:coord}       
\end{figure}

Once constructed, the equivalent Laplacian matrix $\mathcal{L}_{eq}$ is next used to identify the coherent groups of synchronous generators in the system. This can be done via Algorithm $1$ with $\mathcal{L}$ replaced by $\mathcal{L}_{eq}$. Let $\tilde{V}$ consists of eigenvectors corresponding to smallest eigenvalues of $M^{-1}\mathcal{L}_{eq}$. The matrix $\tilde{V}_{r1}$ is constructed by aggregating the rows corresponding to reference machines obtained using Gaussian elimination. Permuting the rows of $\tilde{V}$ we will have
$
 \tilde{V}_r= [\tilde{V}_{r1}^T \;\; \tilde{V}_{r2}^T]^T.
$
Then the row vectors of $ \tilde{V}_{r1}$ are used as unit coordinate vectors in the new coordinate system using the transformation
\begin{equation}
\label{equivalent_transformation2}
 \begin{bmatrix} \tilde{V}_{r1}\\ \tilde{V}_{r2}  \end{bmatrix} \times \tilde{V}_{r1}^{-1} = \begin{bmatrix} I\\\tilde{L}  \end{bmatrix} = \tilde{V}_{L}.
\end{equation} 
Depending on the difference between $\mathcal{L}_0$ and $\mathcal{L}_{eq}$, the matrix $\tilde{L}$ may now produce a different clustering structure than the wind-less system. 
Note that even if the change in the individual entries of the $\mathcal{L}_{eq}$ from $\mathcal{L}_0$ may not be significant, its slow eigenspace can still be notably perturbed so that the coherent groupings before and after wind penetration are different, shown in the following simple example.
Let $\alpha_i, \tilde{\alpha}_{i}, (i =1,\dots,n)$ be the row vectors of the nominal and perturbed eigen-spaces $V_L = [I, L^T]^T$ and $\tilde{V}_L$ following from $\mathcal{L}_0$ and $\mathcal{L}_{eq}$, respectively. From \cite{chowbook} it follows that these perturbed row vectors will also lie on the hyperplane
$
\sum_{j=1}^{r} \tilde{\alpha}_{ij} = 1 , \forall i=1,..,n,
$
just like the nominal row vectors. Fig. \ref{fig:coord} shows the visualization of this hyperplane for a $2-$area $4$-machine system in $2$-dimensional coordinates where $\alpha_i$ and $\tilde{\alpha}_{i}$ are plotted for $i=1,2$. The tip of both row vectors will remain on the same hyperplane as shown in the figure, but their positions may shift depending on $\gamma$ and the wind bus location, which, in turn, can change the coherent grouping. 
\vspace{-.4 cm}

\vspace{-.3 cm}
\section{Simulation Results}

We perform simulations on a $16$-machine, $68$-bus IEEE benchmark power system model - first with a single wind farm and then with three wind farms. This is a simplified model of interconnected New York and New England power systems. 
Simulations have been performed in Matlab $2016a$ platform with intel(R) core(TM)-$i7-2600$ CPU$@3.40$ GHz processor. The total active load is $17.6$ GW. 
Each wind turbine-generator unit is rated at $1.76$ MW and wind parameters are taken from \cite{Souvik_ts}. The internal PI controllers in the active and reactive power loops of the DFIGs are tuned to achieve stable operation.
\vspace{-.5 cm}
\subsection{Nominal system without wind plant}
\begin{figure}[H]
                \centering
                \includegraphics[width=1\linewidth,height= 3 cm]{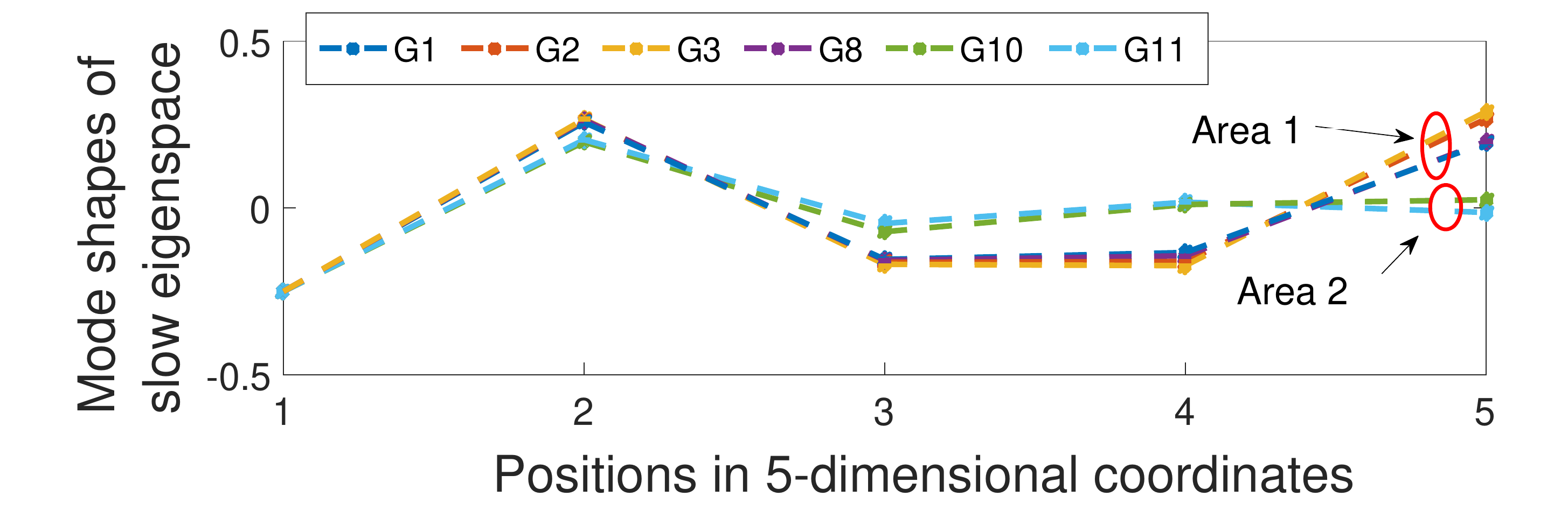}
                \caption{Row vectors of slow subspace of the nominal system}
                \label{fig:rownom}
\end{figure}

The slow oscillation modes of the system without any wind plant and the corresponding coherency structure using Algorithm $1$ are shown in Table-\ref{table:eig_nowind}. The areas are marked in the system diagram in Fig. \ref{fig:coh_nom}. The orientation of row vectors of the slow eigenspace are shown in Fig. \ref{fig:rownom}. The coherency algorithm assigns machines $5$ and $13$ as reference machines for Area $1$ and Area $2$, respectively.

\begin{table}[H]
\label{table:nominal}
\caption{Inter-area modes and areas of the nominal system} 
\centering 
\begin{tabular}{c |c } 
\hline\hline 
Slow modes & Frequency in Hz\\  

\hline 
$-0.1321 \pm j1.877$ & 0.299  \\ 
$-0.1449 \pm j2.843$ & 0.453 \\
$-0.1434 \pm j3.763$ & 0.599 \\
$-0.1772 \pm j3.910$ & 0.622 \\                  
\hline 
\end{tabular}
\quad
\begin{tabular}{|l|l|l|l|l|}
\hline
A1 & \multicolumn{4}{l|}{5,1,2,3,4,6,7,8,9} \\ \hline
A2 & \multicolumn{4}{l|}{13,10,11,12}       \\ \hline
A3 & \multicolumn{4}{l|}{14}                \\ \hline
A4 & \multicolumn{4}{l|}{15}                \\ \hline
A5 & \multicolumn{4}{l|}{16}                \\ \hline
\end{tabular}
\label{table:eig_nowind} 
\end{table}
\vspace{-.5 cm}
\vspace{-1 cm}
\subsection{Single wind plant connected to the grid}
Several subcases are considered with the wind plant connected to buses $66, 37, 32$, and $38$. Bus location $66$ is considered as it is away from loads and generation zones in Area $1$. Bus $37$ is tested because of its close proximity to the loads in this area. Total connected load in Area $1$ is more than the total generation. Bus locations $32$ and $38$ belong to Area $2$, where total connected load is less than the generation. Here we increase wind penetration to higher values in order to test the change in coherency behavior. 
\vspace{-.4 cm}
\hspace{.5 cm} \begin{figure}[H]
\centering
 \includegraphics[width=\linewidth,height= 3 cm]{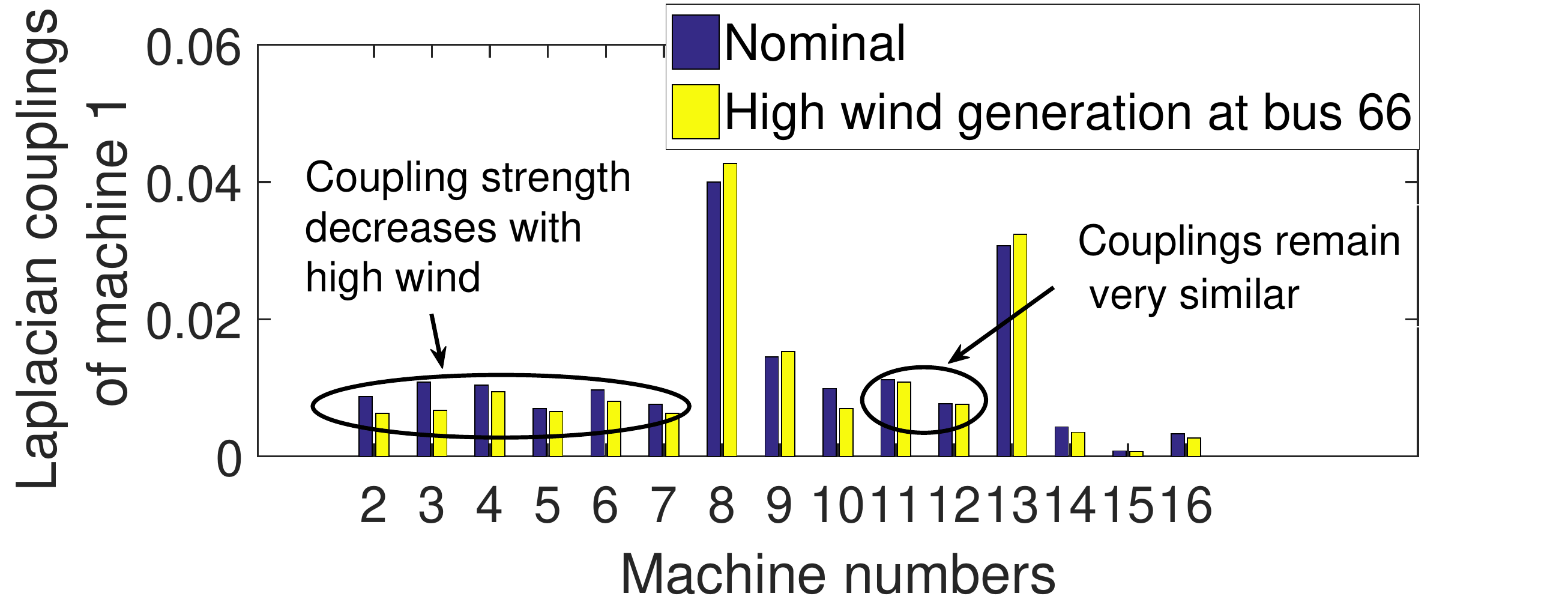}
\caption{Change in the entries of $\mathcal{L}_{eq}$ vs. $\mathcal{L}_0$ for machine $1$ with wind injection at bus $66$}
\label{fig:lap_bar}
\end{figure}

\begin{figure*}
        \begin{subfigure}[b]{0.5\textwidth}
                \centering
                \includegraphics[width=.93\textwidth,height=9\textheight,keepaspectratio,trim={0 0 0 0},clip]{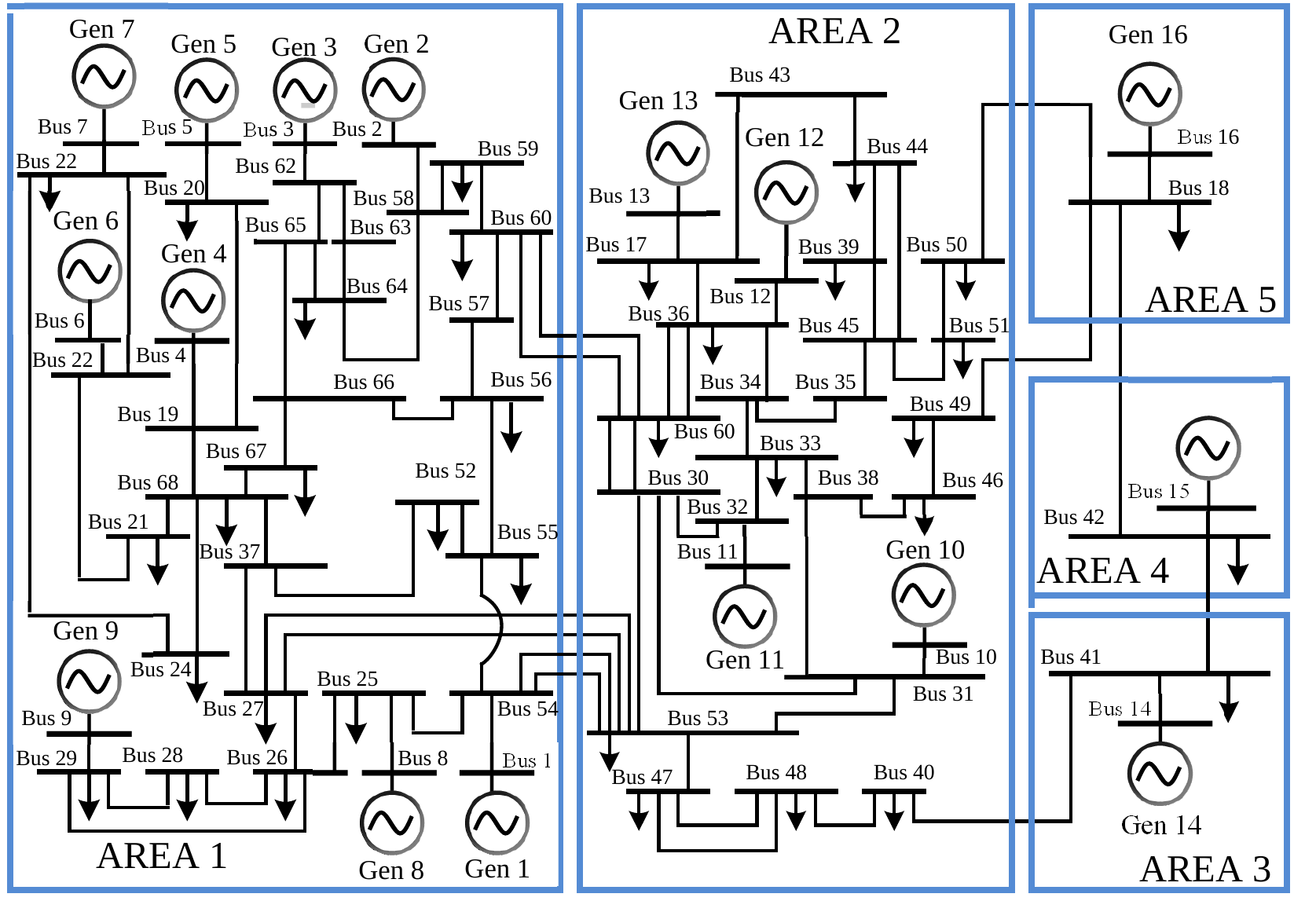}
                \caption{Nominal system with five coherent areas}
                \label{fig:coh_nom}
        \end{subfigure}
       \qquad
             \begin{subfigure}[b]{0.5\textwidth}
                \centering
                \includegraphics[width=.93\textwidth,height=.75\textheight,keepaspectratio, trim={0 0 0 0},clip]{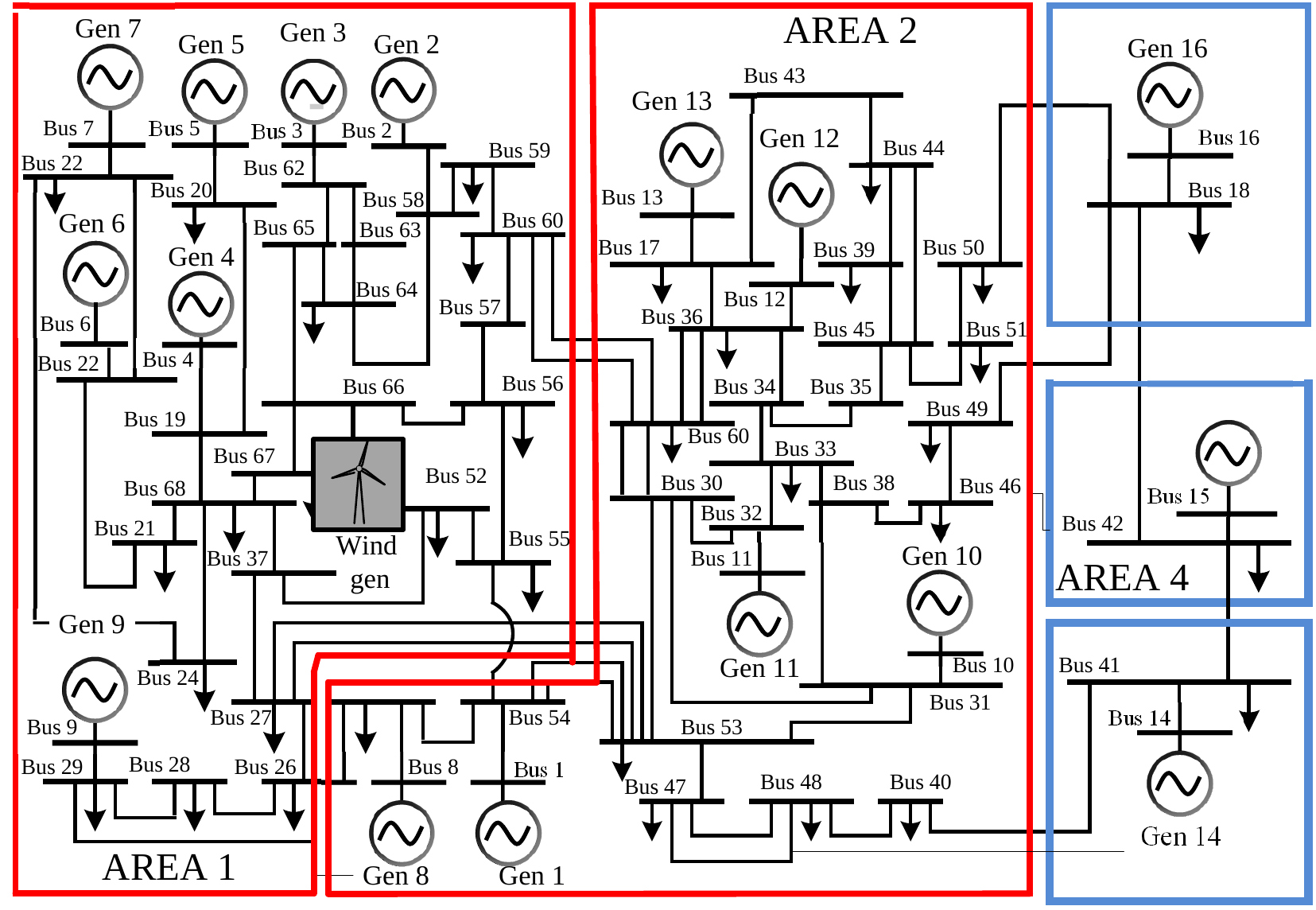}
                \caption{Modified coherent grouping}
                \label{fig:coh_change}
        \end{subfigure}
        
\caption{\small{Change in coherency grouping when $1144$ MW of  wind power is injected at bus $66$ }}
\label{fig:coherent}
\end{figure*}

\vspace{-.4 cm}
\begin{figure}[H]
                \centering
                \includegraphics[width=1\linewidth,height= 3 cm]{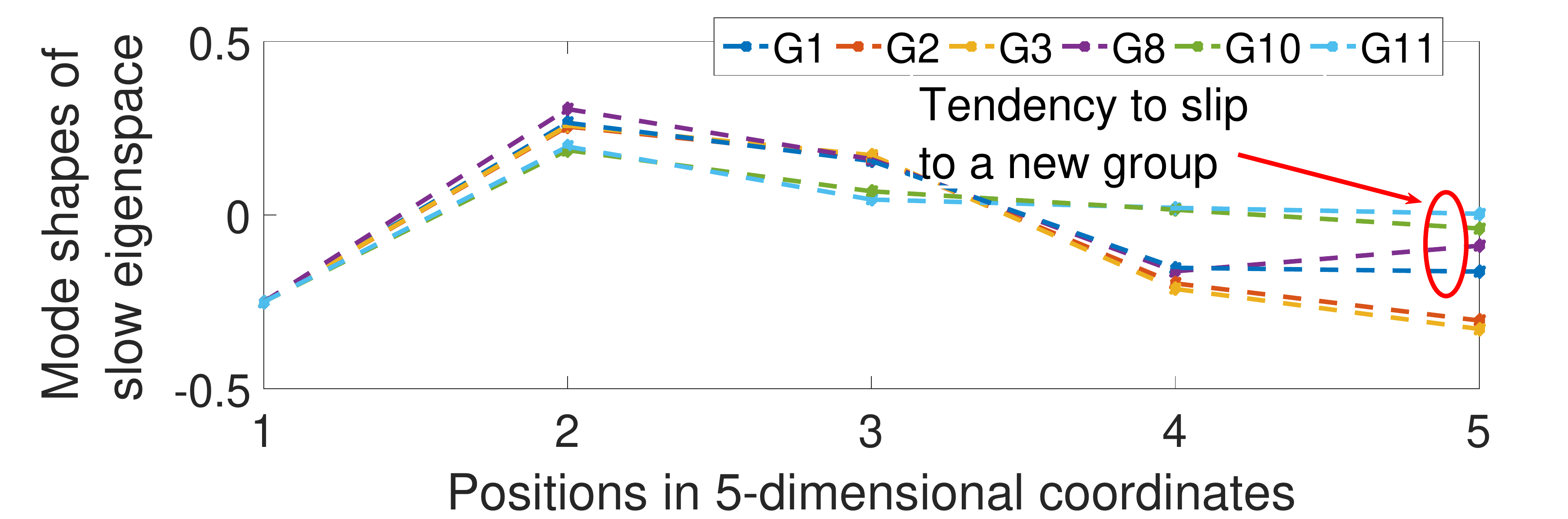}
                \caption{Row vectors of slow eigenspace, wind at bus $66$}
                \label{fig:row}
                \vspace{-.4 cm}
\end{figure}

\begin{figure}[H]
    \centering
    \begin{minipage}{0.24\textwidth}
        \centering
        \includegraphics[width=1.0\linewidth,height=3.5 cm]{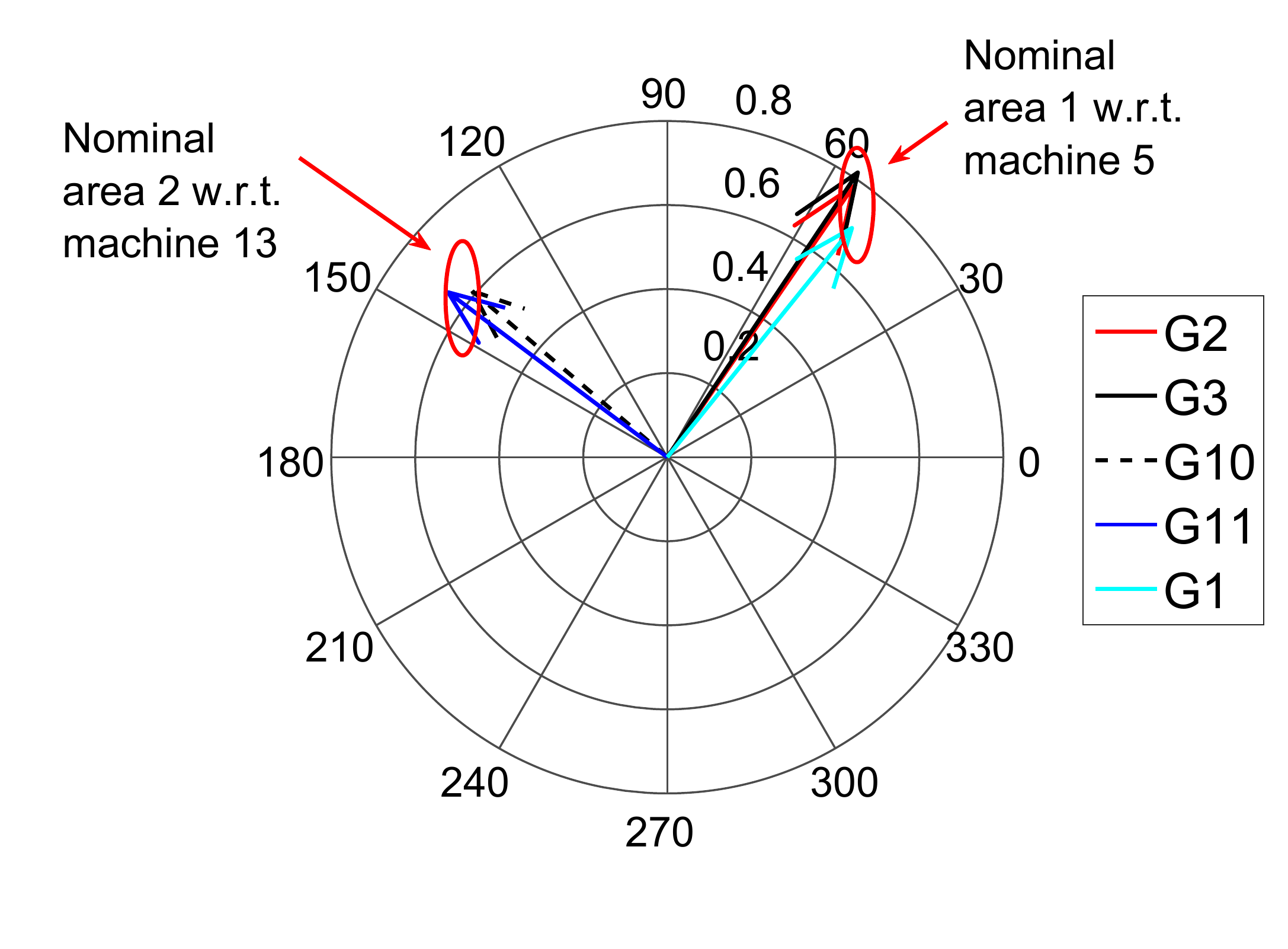}\\
         \hspace{.3 cm} 8.a Nominal system
    \end{minipage}
     \begin{minipage}{0.23\textwidth}
        \centering
        \includegraphics[width=1.05\linewidth,height= 3.5 cm]{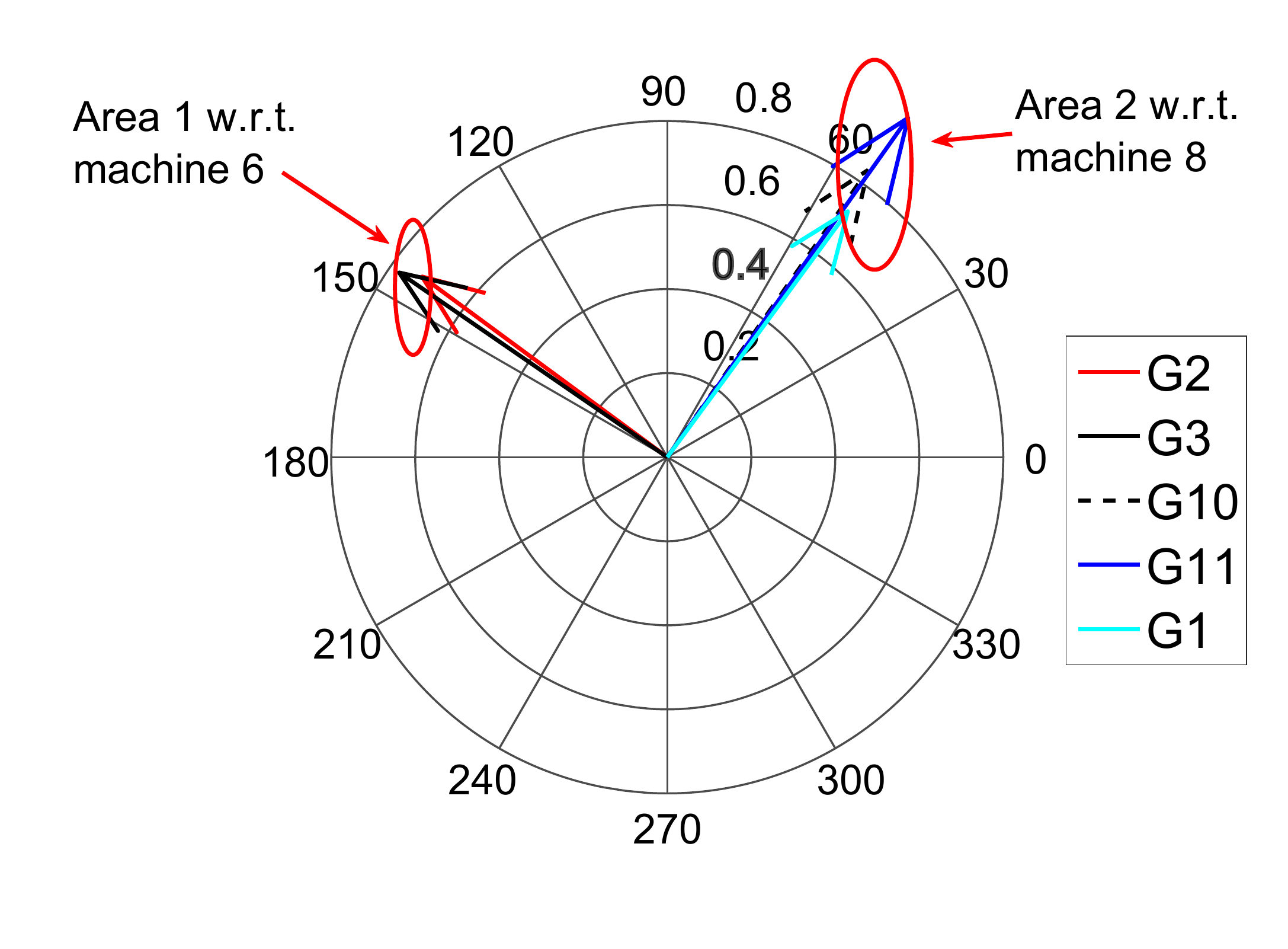}\\
        \hspace{.3 cm} 8.b System with wind at bus $66$
    \end{minipage}
    \caption{Compass plots of selected rows of $\tilde{V}_L$}
    \label{fig:compass}
    \vspace{-.8 cm}
    \end{figure}

We first present the case where the wind plant is located at bus $66$ with $\gamma = 650$ ($1144$ MW). The blue and yellow stems in Fig. \ref{fig:lap_bar} denote the values of the entries of $\mathcal{L}_0$ and $\mathcal{L}_{eq}$, showing how the row entries of the Laplacian corresponding to generator $1$ change before and after wind penetration.
The wind-integrated model has four slow modes as shown in Table-\ref{table:eig_wind66_650}. We construct $\mathcal{L}_{eq}$ and apply algorithm 1 to identify the five slow coherent areas arising from these four slow modes. These areas and the indices of the synchronous generators in each area are also shown in Table-\ref{table:eig_wind66_650}. Comparing tables I and II, it can be clearly seen that the wind injection forces generators $1$ and $8$ to move from Area 1 to Area 2. The other generators remain in their respective areas. Fig. \ref{fig:coherent} shows the system diagram with modified clustering before and after wind injection. Fig. \ref{fig:row} shows the row vectors of the perturbed slow eigen-space $\tilde{V}$ while Fig. \ref{fig:compass} shows the compass plots for the selected row vectors of the transformed eigen-space $V_L$ of the nominal system (Fig. 8a) versus $\tilde{V}_L$ of the perturbed system (Fig. 8b). Both figures testify to the tendency of generators $1$ and $8$ to move from Area 1 to Area 2. 

\vspace{-.33 cm}
\begin{table}[H]
\caption{Inter-area modes and areas (bus 66, $\gamma=650$)} 
\centering 
\begin{tabular}{c |c } 
\hline\hline 
Slow modes & Frequency in Hz\\  

\hline 
$-0.129 \pm j1.761$ & 0.280  \\ 
$-0.145 \pm j2.825$ & 0.449 \\
$-0.142 \pm j3.752$ & 0.597 \\
$-0.1436 \pm j3.844$ & 0.612 \\                  
\hline 
\end{tabular}
\quad
\begin{tabular}{|l|l|l|l|l|}
\hline
A1 & \multicolumn{4}{l|}{6,2,3,4,5,7,9}   \\ \hline
A2 & \multicolumn{4}{l|}{8,10,11,12,13,1} \\ \hline
A3 & \multicolumn{4}{l|}{14}              \\ \hline
A4 & \multicolumn{4}{l|}{15}              \\ \hline
A5 & \multicolumn{4}{l|}{16}              \\ \hline

\end{tabular}
\label{table:eig_wind66_650} 
\end{table}
\vspace{-.3 cm}
Next we consider the case when the wind plant is located at bus $37$ with $\gamma = 700$ ($1232$ MW). Table-\ref{table:eig_wind37_700} shows the slow modes and the corresponding coherent clusters. Compared to the previous case the generator indices in the respective areas do not change; however, the reference machine of Area $1$ now changes from $6$ to $5$.
\vspace{-.3 cm}
\begin{table}[H]
\caption{Inter-area modes and areas (bus $37$, $\gamma$ = $700$)} 
\centering 
\begin{tabular}{c| c } 
\hline\hline 
Slow modes & Frequency in Hz\\  

\hline 
$-0.1299 \pm j1.82$ & 0.289  \\ 
$-0.1447 \pm j2.83$ & 0.450 \\
$-0.1416 \pm j3.756$ & 0.598 \\
$-0.1698 \pm j3.95$ & 0.629 \\                  
\hline 
\end{tabular}
\quad
\begin{tabular}{|l|l|l|l|l|}
\hline
A1 & \multicolumn{4}{l|}{5,2,3,4,6,7,9}   \\ \hline
A2 & \multicolumn{4}{l|}{8,10,11,12,13,1} \\ \hline
A3 & \multicolumn{4}{l|}{14}              \\ \hline
A4 & \multicolumn{4}{l|}{15}              \\ \hline
A5 & \multicolumn{4}{l|}{16}              \\ \hline
\end{tabular}
\label{table:eig_wind37_700} 
\end{table} 

\vspace{-.35 cm}
We further validate our results for this case using the model-free PCA technique. The comparison is done only for the validation of our derivations, and not intended for comparison between model-driven and data-driven methods. PCA is carried out on the rotor angle measurements of all $16$ generators for $100$ seconds with a sampling time of $0.01$ s, following a unit step change in the mechanical power of machine $1$. Fig. 9 shows the first three principal components with the maximum  variance in the data set. When the wind plant is connected to bus $37$ with $\gamma=700$ the plot clearly shows that the positions of generators $1$ and $8$ move more towards generators $11$ and $12$. The same is predicted by the coherency algorithm using $\mathcal{L}_{eq}$.
\begin{figure}[H]
\centering
\includegraphics[width=\linewidth,height= 3.2 cm,trim={4 4 4 4},clip]{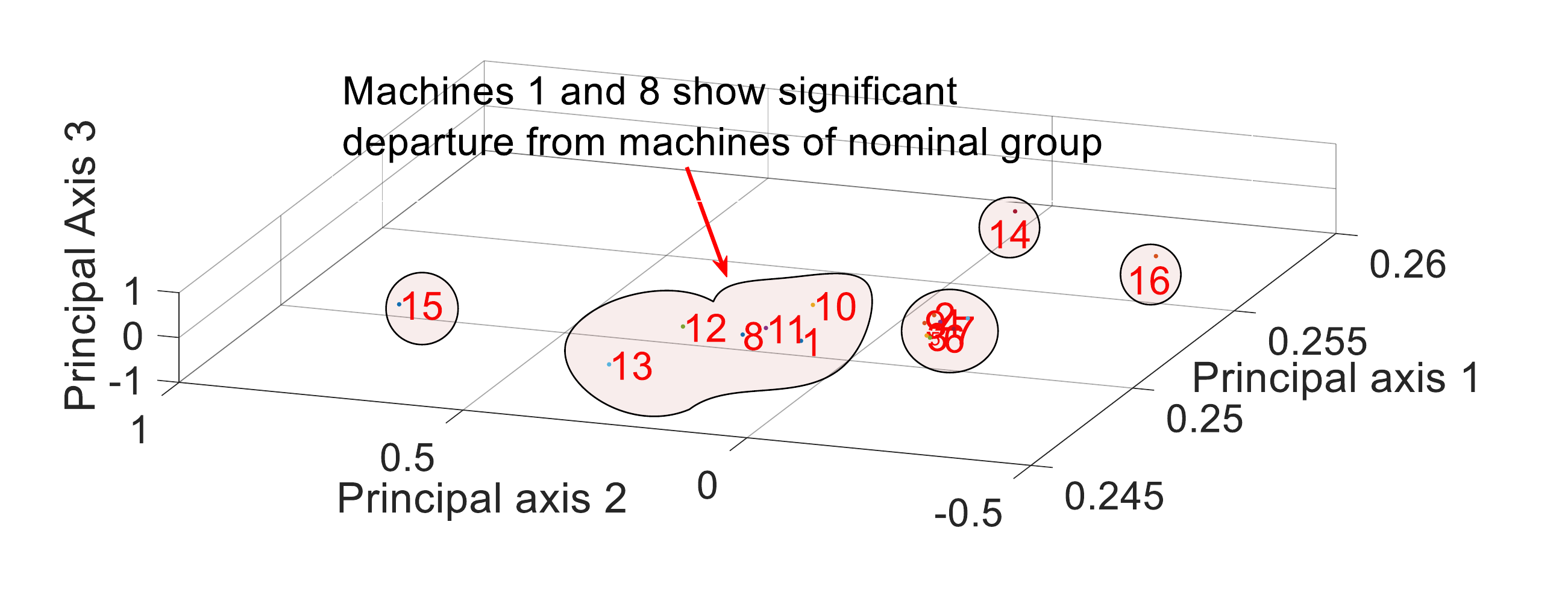}
\caption{\small{PCA coefficients from generator angle data (bus-$37$,$\gamma =700$)} }
\label{fig:pca37}
\vspace{-.3 cm}
\end{figure}



The third case considers the wind plant connected at bus $32$ in Area $2$. For this case even when $\gamma = 700$, the coherency structure was not found to change from the nominal. Similar observation is made when the wind plant is located at bus $38$. This indicates that Area $1$ is more prone to perturbation in coherency than Area $2$. This can be a useful message for transmission planners for deciding the location of wind installations without disturbing the coherency of their grid. 
\vspace{-.5 cm}
\subsection{Wind plants at buses $32,66$ and $57$}
\begin{figure}[H]
\centering
\includegraphics[width=1\linewidth,height= 2.7 cm]{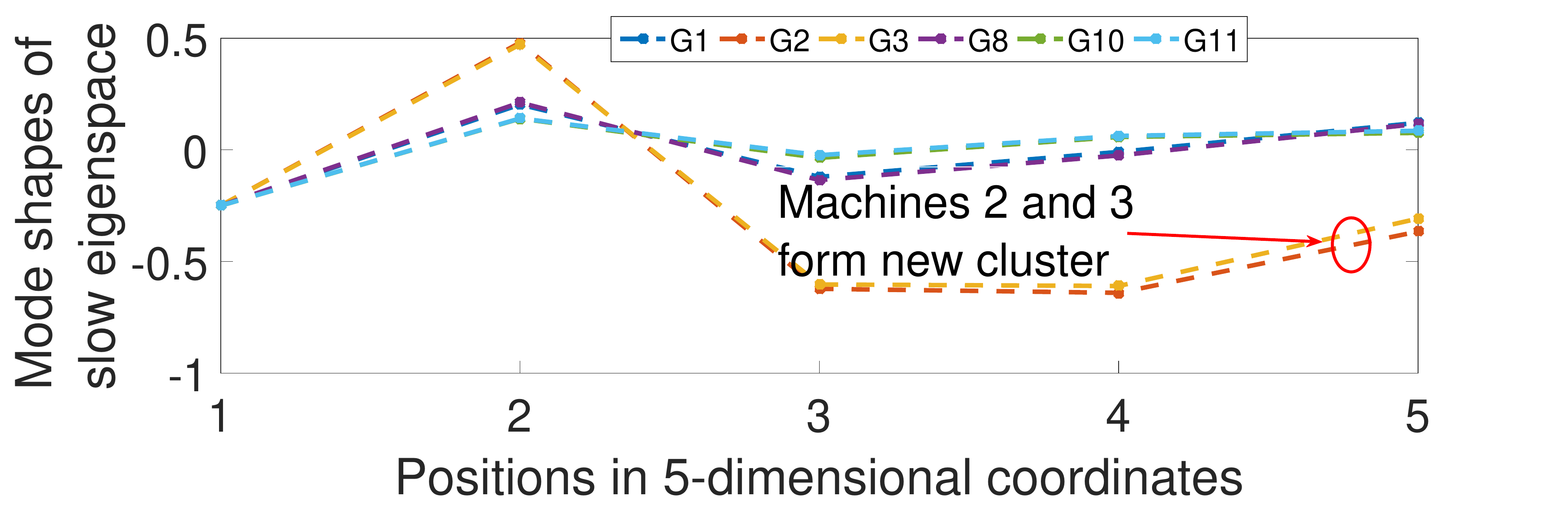}
\caption{\small{Selected row vectors of $\tilde{V}$ for the $3$-wind farm scenario } }
\label{fig:row326657}
\includegraphics[width=\linewidth,height= 3 cm]{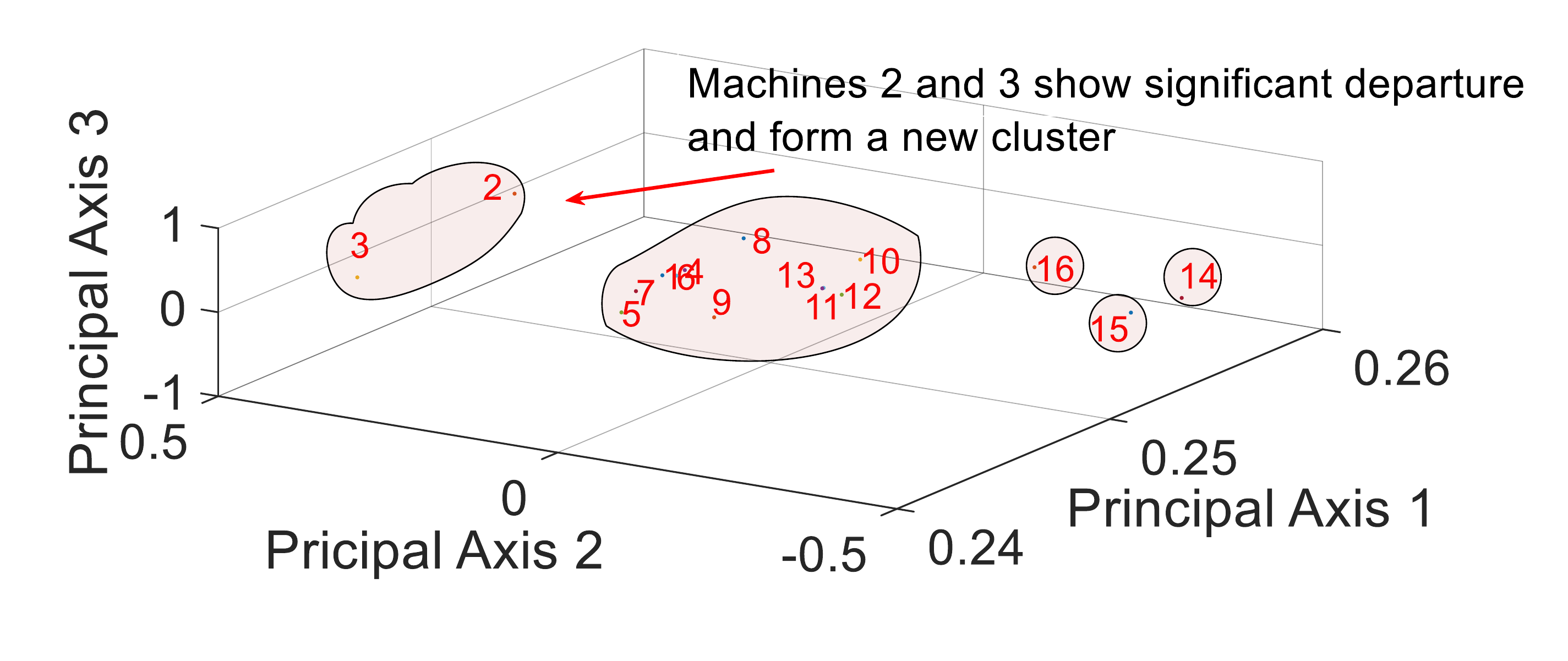}
\caption{\small{PCA coefficients from generator angle data for the $3$-wind farm scenario} }
\label{fig:pca326657}
\end{figure}

We next consider the IEEE $68$-bus system with three wind farms located at buses $32,66$ and $57$ with penetration levels  $\gamma_1 = 200$, $\gamma_2 = 250$, $\gamma_3 = 200$, respectively. The corresponding $\mathcal{L}_{eq}$ is constructed and Algorithm 1 is applied. The resulting clustering structure is shown in Table-\ref{table:eig_wind3}. The table shows that with wind installed at these three locations, Area 1 now shrinks to only two generators - namely, generators $2$ and $3$. Area $2$, on the other hand, now expands to a much bigger geographical area covering a total of $11$ generators. Areas $3$ through $5$, however, remain unaffected. The row vectors of $\tilde{V}$ are plotted in Fig. \ref{fig:row326657} indicating the same result. The result is also validated using PCA, as shown in Fig. 11, where generators $2$ and $3$ depart from their nominal dynamic signatures forming an area between just the two of them. 
\begin{table}[H]
\caption{Inter-area modes and areas with $3$ wind plants} 
\centering 
\begin{tabular}{c |c } 
\hline\hline 
Slow modes & Frequency in Hz\\  

\hline 
$-0.1363 \pm j2.244$ & 0.3571  \\ 
$-0.1666 \pm j2.65$ & 0.4219 \\
$-0.1348 \pm j3.118$ & 0.4964 \\
$-0.1482 \pm j3.986$ & 0.6342 \\                  
\hline 
\end{tabular}
\quad
\begin{tabular}{|l|l|}
\hline
A1 & 2,3                                                                  \\ \hline
A2 & \begin{tabular}[c]{@{}l@{}}13,1,4,5,6,7,\\ 8,9,10,11,12\end{tabular} \\ \hline
A3 & 14                                                                   \\ \hline
A4 & 15                                                                   \\ \hline
A5 & 16                                                                   \\ \hline
\end{tabular}
\label{table:eig_wind3} 
\end{table}

\vspace{-.6 cm}


\vspace{-1 cm}
\section{Conclusion}
A mathematical analysis of the perturbation in coherency of synchronous generators due to wind integration is presented in this paper. The dynamic interaction between the generators in a wind-integrated system is captured by an \textit{equivalent} Laplacian matrix. Depending on the amount of wind injection and placement of wind plants the slow eigenspace of the \textit{equivalent} Laplacian matrix may change, thereby changing the coherent groupings. Results are validated using the IEEE $68$-bus system with single and multiple wind farms. The results can be useful for transmission planners in deciding potential locations of wind installations, and also for readjusting wide-area control gains in case the wind injection changes the coherent groupings. 

\bibliographystyle{IEEEtran}
\vspace{-.4 cm}
\bibliography{ref_pes}
\end{document}